\documentclass[12pt]{iopart}
\usepackage{graphicx}
\usepackage{bm}

\newcommand{\simle}
{\raisebox{-0.75ex}[-1.5ex]{$\;\stackrel{<}{\sim}\;$}}
\newcommand{\simge}
{\raisebox{-0.75ex}[-1.5ex]{$\;\stackrel{>}{\sim}\;$}}
\def\rd{{\partial}}
\def\sg{{\sigma}}
\def\eps{{\varepsilon}}
\newcommand{\sxy}{\sigma_{\rm SH}}
\def\bk{{ {\bm k} }}

\def\g{{\gamma}}

\newcommand{\eq}{eqnarray}
\newcommand{\f}{\frac}
\def\nn{{\nonumber}}
\newcommand{\aH}{\alpha_{\rm SH}}

\begin{document}

\title{Giant Extrinsic Spin Hall Effect
due to Rare-Earth Impurities }

\author{
T. {\sc Tanaka} and H. {\sc Kontani}
}

\address{
Department of Physics, Nagoya University,
Furo-cho, Nagoya 464-8602, Japan. 
}
\ead{
takuro@slab.phys.nagoya-u.ac.jp
}
\begin{abstract}
We investigate the extrinsic spin Hall effect in the electron gas model due to
magnetic impurities, by focusing on Ce- and Yb-impurities.
In the dilute limit, the skew scattering term dominates the side jump term.
For Ce-impurities, the spin Hall angle $\alpha_{\rm SH}$ due to skew scattering is given by $-8\pi \delta_2/7$, where $\delta_2 \ (\ll 1)$ is the phase shift for $d \ (l=2)$ partial wave.
Since $\alpha_{\rm SH}$ reaches $O(10^{-1})$ if $\delta_2 \simge 0.03$,
considerably large spin Hall effect is expected to emerge in metals with
rare-earth impurities.
The present study provides the highly efficient way to generate a  
spin current.

\end{abstract}

\maketitle

\section{Introduction}

The spin Hall effect (SHE) has been attracting a great deal of interest 
due to its fundamental as well as technological importance. 
The intrinsic SHE originates from the Berry phase of the
multiband Bloch function, so that the spin Hall conductivity (SHC)
takes a material-specific value that is independent of impurity scattering
in the low resistivity regime
\cite{Sinova-SHE, Murakami-SHE}.
In transition metals, the intrinsic SHC takes a
considerably large value \cite{Saitoh, Kimura},
because of the ``effective Aharonov-Bohm (AB) phase" induced by
$d$-orbital degrees of freedom and the atomic spin-orbit interaction (SOI)
\cite{Kontani-Ru, Kontani-Pt, Tanaka-4d5d, Kontani-OHE}.
Intrinsic SHE in Sr$_2$RuO$_4$ had been calculated in ref. \cite{Kontani-Ru}
based on the $t_{2g}$ tight-binding model.
By the same mechanism, SHE in transition metals had been studied 
based on the realistic multiorbital tight-binding model
in ref. \cite{Tanaka-4d5d} and the first-principles calculations \cite{Guo-Pt}. 
The obtained SHCs agree well with 
recent experimental observations, in both the magnitude and sign \cite{Kontani-OHE}.

In addition to the intrinsic mechanism, SHE is also induced by
impurity scattering with the aid of SOI;
this phenomenon is known as the extrinsic SHE \cite{Dyakonov,Hirsh,Takahashi}.
It consists of a skew scattering term ($\sigma_{\rm SH}^{\rm ss}\propto
\rho^{-1}$) and a side jump term ($\sigma_{\rm SH}^{\rm sj}\propto \rho^0$); 
This fact suggests that considerable large SHC can be realized due to the 
skew scattering term in the very low resistivity metals, beyond the 
material-specific intrinsic SHC.
For this reason, study of the extrinsic SHE is significant for realizing excellent 
spintronics device to produce or detect the spin current.


The extrinsic SHE due to nonmagnetic impurities or vacancies
had been studied in refs. \cite{Dyakonov,Hirsh,Takahashi},
by developing the theory of the extrinsic anomalous Hall effect (AHE)
\cite{Smit, Berger, Bruno}.
The obtained SHC is,
however, very small unless the enhancement factor for the SOI
due to the multiband effect is taken into consideration.
For this reason, it is not easy to estimate the magnitudes of 
$\sigma_{\rm SH}^{\rm ss}$ and $\sigma_{\rm SH}^{\rm sj}$
due to nonmagnetic impurities.
In transition metals, there is no apparent experimental evidence
for a large extrinsic SHE due to natural randomness. 

Dilute magnetic impurities such as Fe and Ce
have been found to induce a large extrinsic AHE in simple metals
\cite{Fert-Jaoul, Fert}.
The Hall angles in AuFe \cite{Fert-Jaoul} and rare-earth doped metals 
\cite{Fert} reach  0.01$\sim$0.001 under high magnetic fields.
Since the SHE has a similar mechanism to that of the AHE,
one might expect that a giant SHE can be realized by magnetic impurities.
However, theoretical analysis of the extrinsic SHE in a simple $s$-electron metal
due to magnetic impurities has not been performed previously.

In this paper, we study the extrinsic SHE based on a single-impurity
Anderson model for Ce and Yb atoms, 
both of which are typical magnetic impurities.
Nonperturbative effect of the large SOI for $f$-electrons
($\sim3000$ K) is correctly accounted for.
Due to skew scattering, the obtained spin Hall angle 
$\displaystyle \aH = \f{j^{\rm s}_{x}}{j^{\rm c}_{y}} \f{2|e|}{\hbar}$, which is the ratio 
between the transverse spin Hall current and the longitudinal electric current, 
exceeds $0.1$ for $\delta_2 \simge 0.03$, where $\delta_2 \ (\ll 1)$ is 
the phase shift for $d \ (l=2)$ partial wave;
It is more than 10 times greater than that in nonmagnetic
metals such as Pt \cite{Kimura}.
Since the spin Hall angle represents the efficiency of the creating spin current,
giant spin Hall current is expected to emerge in simple metals
by introducing dilute magnetic impurities.
Moreover, the extrinsic SHC can take a large value
well above the Kondo temperature $T_{\rm K}$, in contrast to 
the anomalous Hall conductivity (AHC).  
The origin of the giant SHE is found to be the large SOI and angular momentum
in the rare-earth impurities, which had not been treated appropriately
in previous studies of the extrinsic SHE.



\section{Model and Hamiltonian}

In this study, we use the single-impurity Anderson model for Ce and Yb atoms.
In a Ce$^{3+}$ ion with a $4f^1$ configuration,
the $J=7/2$ level is about 3000 K higher than the $J=5/2$ level
due to the strong atomic SOI \cite{Hewson}.
Therefore, we consider only $J=5/2$ states for the Ce-impurity.
In the same way, we consider only $J=7/2$ states in Yb$^{3+}$ ion
with $4f^{13}$ configuration.
Here, we introduce the following single-impurity Anderson model
for Ce$/$Yb atoms with both $d$- and $f$-orbitals \cite{Fert}:
\begin{eqnarray}
H&=& \sum_{\bk,\sg} \eps_{\bk} c^{\dagger}_{\bk \sg} c_{\bk \sg}
+ \sum_{\bk\sg m} E^d d^{\dagger}_{\sg m} d_{\sg m} + \sum_{\bk M} E^f f^{\dagger}_M f_M \nn \\
&+&\sum_{\bk\sg m} \left\{ V^{d}_{\bk m} c^{\dagger}_{\bk \sg} d_{\sg m} + {\rm h.c.} \right\} 
+ \sum_{\bk\sg M} \left\{ V^{f}_{\bk M \sg} c^{\dagger}_{\bk \sg} f_M + \rm {h.c.} \right\} \nn \\
&+& \f{U^f}{2} \sum_{M \neq M'} n^f_M n^f_{M'},  \label{eq:Ham}
\end{eqnarray}
where we have employed the electron (hole) picture for the Ce (Yb) impurity.
$c^{\dagger}_{\bk \sg}$ is the creation operator of a conduction 
electron with spin $\sg=\pm 1$. 
$f^{\dagger}_{M}$ is the creation operator of a $f$-electron with 
total angular momentum $J=5/2 \ (7/2)$ and $z$-component 
$M \ (-J \leq M \leq J )$ for Ce$^{3+}$ (Yb$^{3+}$).  
$d^{\dagger}_{\sg m}$ is the creation operator of a $d$-electron with 
angular momentum $m(-2 \leq m \leq 2 )$ \cite{comment}.
%
%
$\eps_{\bk}=k^2/2m$ is the energy for the conduction-electrons, and $E^f$ ($E^d$) is the localized $f(d)$-level energy. 
$V^{f}_{\bk M \sg}$  and $V^{d}_{\bk m}$ are the mixing potentials, which are given 
by
\begin{\eq}  
V^{f}_{\bk M \sg} &=& \sqrt{4\pi} V_f \sum_{m} a^M_{m\sg} Y^m_3 (\hat \bk),\\
V^{d}_{\bk m} &=& \sqrt{4\pi} V_d Y^m_2 (\hat \bk),
\end{\eq}
where $a^M_{m \sg}$ is the Clebsch-Gordan (C-G) coefficient and 
$Y^m_l(\hat \bk)$ is the spherical harmonic function.
We will show that the phase factor in $Y_l^m(\hat \bk)$ and 
$a^M_{m\sg}$ are indispensable to realize the SHE. 
For $J=5/2$, 
$a^M_{m\sg} = -\sg \left\{ \left( 7/2-M\sg \right)/7  \right\}^{1/2} \delta_{m,M-\sg/2}$ \cite{YY, Kontani94}.
%
Note that the SOI was neglected in the study of the AHE in ref. \cite{Fert}. 
In the present study, we neglect the crystalline electric field of the $f$-orbitals, 
since it is small due to the small radius of the $f$-orbital wave function 
\cite{Fert}.
We put $\hbar = 1$ hereafter.

To discuss the scattering problem, it is useful to
derive an effective Hamiltonian for the conduction electrons
by integrating out the 
$f$ and $d$ electrons in eq. (\ref{eq:Ham}).
The obtained Hamiltonian is given by \cite{Fert}
\begin{\eq}
H_c&=& \sum_{\bk \sg} \eps_{\bk} c^{\dagger}_{\bk \sg}c_{\bk \sg} 
+ \sum_{\bk,\bk',\sg} J^d_{\bk,\bk'} c^{\dagger}_{\bk \sg} c_{\bk' \sg} 
+ \sum_{\bk \bk'\sg,\sg'} J^f_{\bk\sg,\bk'\sg'} c^{\dagger}_{\bk\sg}c_{\bk' \sg'},
\label{eq:Ham-re}
\end{\eq}
where
\begin{\eq}
J^d_{\bk,\bk'} &=& \f{1}{\mu-E^d}\sum_{m} V^d_{\bk m} V^{d \ast}_{\bk' m} \nn \\
&=& 4\pi J_d \sum_{m} Y^m_2(\hat \bk) \left[ Y^m_2(\hat \bk') \right]^{\ast},  \\ 
J^f_{\bk\sg,k'\sg'} &=& \f{1}{\mu-\tilde E^f}\sum_{M} V^f_{\bk M\sg} V^{f \ast}_{\bk' M\sg'} \nn \\
&=&4\pi J_f \sum_{M m m'} a^M_{m\sg} a^M_{m' \sg'} Y^m_3 (\hat \bk) \left[ Y^{m'}_3(\hat \bk') \right]^{\ast},
\end{\eq}
and $ J_d \equiv |V_d|^2/(\mu-E^d)$ and $ J_f \equiv |V_f|^2/(\mu-\tilde E^f)$.
Here, $\tilde E^f = E^f + \rm {Re}\Sigma^f$; $\Sigma^f$ is the $f$-electron 
self-energy due to the Coulomb interaction $U^f$ \cite{YY}.
According to the scaling theory \cite{Haldane}, $\tilde E^f $ approaches the 
Fermi level as the temperature decreases due to the Kondo effect:
$|J_f|$ is strongly enhanced near $T_{\rm{K}}$, and below $T_{\rm K}$,
$J_f N(0) \gg 1$ due to strong resonant scattering,
where $N(0)=m k_{F}/2\pi^2$ is the density of state of the conduction band per spin. 
We also assume that $J_d N(0) (=-\tan\delta_2/\pi) \ll 1$ 
since $|\mu - E^d | \sim O(1 {\rm eV})$ \cite{Fert}. 

\section{Calculations}
\subsection{$T$-matrix due to $c$-$f$ Resonant Scattering}

\begin{figure}[!htb]
\includegraphics[width=1.0\linewidth]{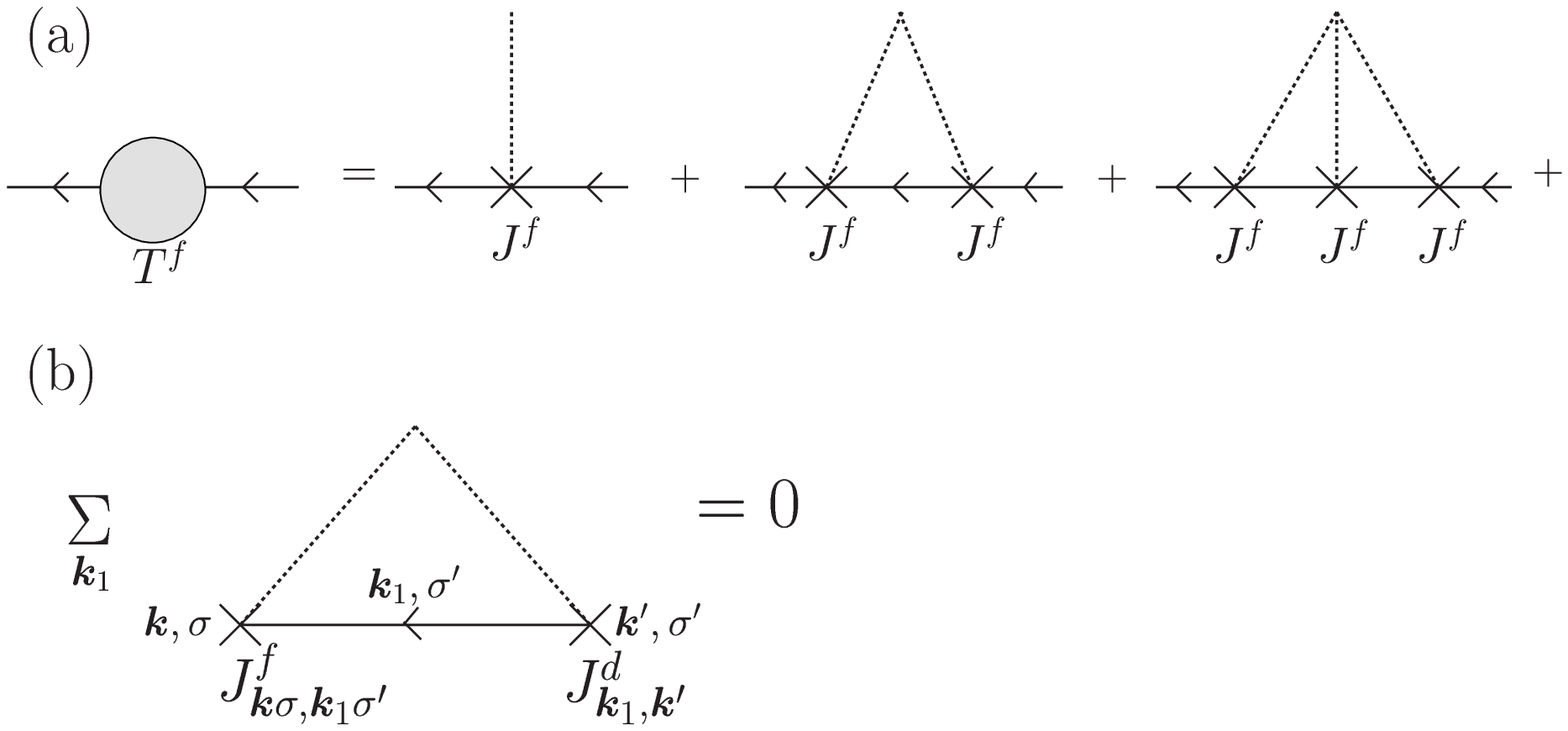}
\caption{\label{fig:tmat} (a) Diagrammatic expression for the $T$-matrix
due to $c$-$f$ resonant scattering. (b) The diagram that vanishes after $\bk_1$
-summation.  
} 
\end{figure}

In this section, we study the scattering problem. 
In the both models given by eq. (\ref{eq:Ham}) and eq. (\ref{eq:Ham-re}),
the $T$-matrix due to the $c$-$f$ resonant scattering is equivalent;
it is given by
\begin{\eq}
T^f_{\bk\sg,\bk'\sg'} = J^f_{\bk\sg,\bk'\sg'} + \frac{1}{N}\sum_{\bk_1,\sg_1} J^f_{\bk\sg,\bk_1\sg_1} G^0_{\bk_1} T^f_{\bk_1\sg_1,\bk'\sg'}, \label{eq:tmat}
\end{\eq}
where its diagrammatic expression is shown in Fig. \ref{fig:tmat} (a),
$N$ is the number of $\bk$-points,
and $G^0_{\bk}=(\eps - \eps_{\bk})^{-1}$.
%
%
We note that the term containing both $J^f_{\bk\sg,\bk'\sg'}$ and $J^d_{\bk,\bk'}$ given in Fig. \ref{fig:tmat} (b) vanishes identically due to 
the orthogonality of spherical harmonic functions.
The solution of eq. (\ref{eq:tmat}) for $\bk=\bk'$ and $\sg=\sg'$ is simply given by
\begin{\eq}
T^f_{\bk\sg,\bk\sg}(\eps)= 3 J_f \f{1}{1-J_f g(\eps)},
\end{\eq}
where we have used the relations $\sum_{M} |V^f_{\bk M \sg}|^2 =3 |V_f|^2$
and $\frac{1}{N} \sum_{\bk \sg} V^f_{\bk M\sg} G_{\bk}^0(\eps) V^{f \ast}_{\bk M'\sg}=|V_f|^2 g(\eps) \delta_{MM'}$.
$g(\eps)=\frac{1}{N} \sum_{\bk} G^0_{\bk}(\eps)$ is the local Green function.
Assuming approximate particle-hole symmetry near $\mu$, we put 
$g^R(0)\equiv g(+i\delta)=-i \pi N(0)$.
%
Then, the quasiparticle damping rate in the $T$-matrix approximation 
is given by
\begin{\eq}
\g_f = -n_{\rm imp} {\rm Im} T^{fR}_{\bk\sg,\bk\sg}(0) = 3 n_{\rm imp} \f{\pi N(0) J_f^2}{1+ (\pi N(0) J_f)^2},  \label{eq:gamma-T}
\end{\eq}
where $n_{\rm imp}$ is the impurity concentration.
Note that eq. (\ref{eq:gamma-T}) is exact for $n_{\rm imp} \ll 1$.



\subsection{Skew Scattering Term \label{SS-term}}

\begin{figure}[!htb]
\includegraphics[width=1.0\linewidth]{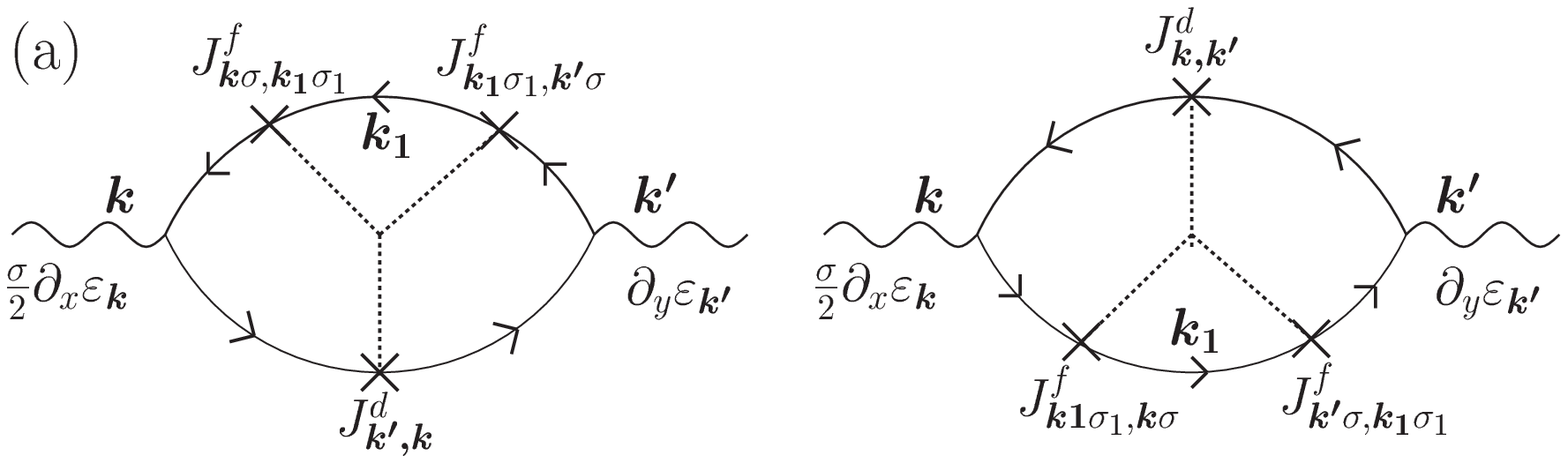}
\includegraphics[width=1.0\linewidth]{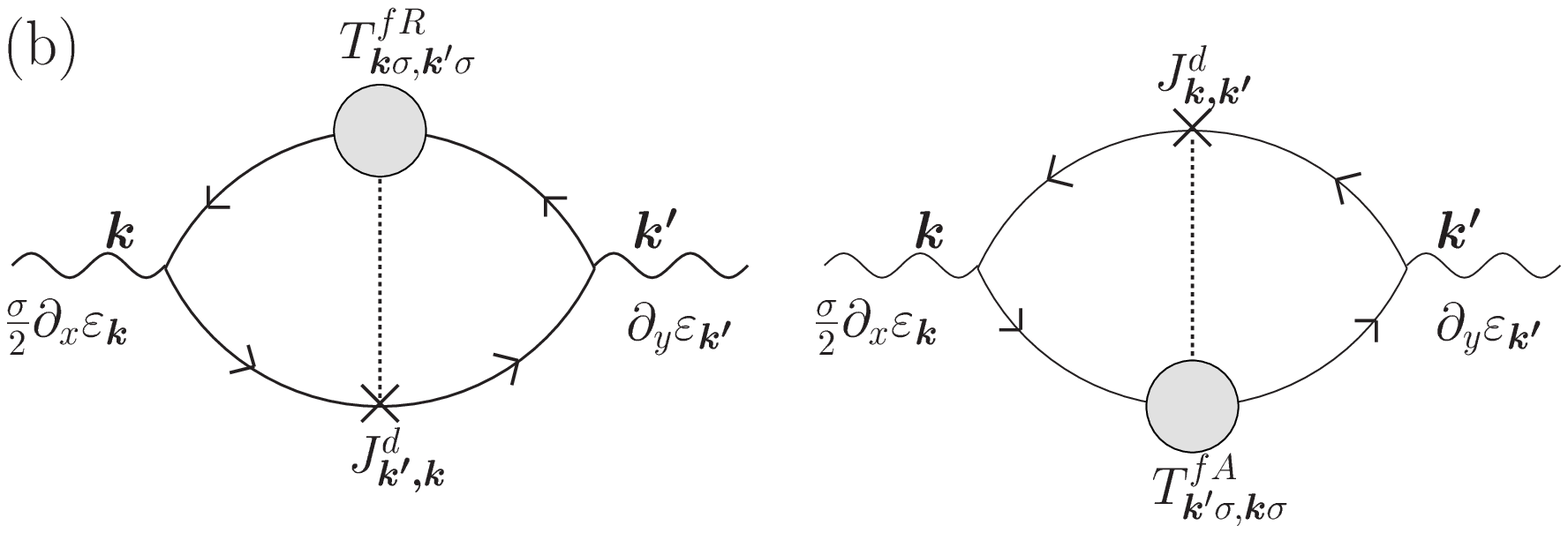}
\caption{\label{fig:ss-born} 
Diagrammatic expressions for SHC induced by
skew scattering (a) within the lowest order (extended Born 
approximation) contribution, and  
(b) in the $T$-matrix approximation with full order diagrams.
} 
\end{figure}

In this section, we study the skew scattering term using linear response theory.
Initially, we consider the case $N(0) J_f \ll 1$ 
and $J_d \ll J_f$, where the Born approximation is valid.
%
In analogy to \cite{Bruno, Fert}, the skew scattering term is given by
\begin{\eq}
\sigma^{\rm ss}_{\rm SH} &=& -\frac{e}{2\pi} n_{\rm imp} \frac{1}{N^2} \sum_{\bm{k,k'} 
\sigma } \frac{\sigma}{2} \frac{\partial 
\varepsilon_{\bm{k}}}{\partial k_x} \frac{\partial \varepsilon_{\bm{k'}}}{\partial k'_y} 
|G^R_{\bm{k}}(0)|^2  \nn \\
&\times& |G^R_{\bm{k'}}(0)|^2 \left\{ T^{f(2)R}_{\bk\sg,\bk'\sg}(0) J^d_{\bm{k',k}} + {\rm c.c} \right\},  \label{eq:ss-born1}  
\end{\eq}
where $-e \ (e>0)$ is the electron charge. 
Its diagrammatic expression is shown in Fig. \ref{fig:ss-born} (a).
%
%
Here,
\begin{\eq}
T^{f(2)R}_{\bk\sg,\bk'\sg}(0) &=& \frac{1}{N} \sum_{\bk_1,\sg_1} J^{f}_{\bk\sg,\bk_1\sg_1} G^{R}_{\bm{k_1}}(0) J^{f}_{\bk_1\sg_1,\bk'\sg} \nn \\
&=& g^R(0) J^f_{\bk\sg,\bk'\sg} J_f 
\end{\eq}
is the second order term of the $T$-matrix;
the first order term in $J^{f}_{\bk\sg,\bk'\sg}$
does not contribute to $\sigma^{\rm ss}_{\rm SH}$
up to the first order term in $J_d$ \cite{Fert}.
%
Note that any diagram that contains the 
part shown in Fig. \ref{fig:tmat} (b) vanishes identically.
The retarded Green function is given by
$G^{R}_{\bk}(0)=(\mu- \eps_{\bk} + i \g)^{-1}$, where $\gamma$ represents 
the quasiparticle damping rate.
Here, we put 
\begin{\eq}
\g=\g_f + \g_0, 
\end{\eq}
where $\g_0$ is the damping rate
due to nonmagnetic scattering, such as $c$-$d$ scattering ($\g_d = 5\pi 
n_{\rm imp}N(0) J^2_d$) and the scattering due to disorders. 
%
The charge current is given by $j^c_{\nu}=-e\rd \eps_{\bk}/\rd k_{\nu}$,
where $\nu=x,y$.
The spin current is then given by $j^s_{\nu} = (\sg/2) \rd \eps_{\bk}/\rd k_{\nu}$.

First, we consider the angular integration in eq. (\ref{eq:ss-born1}),
which is given by
\begin{\eq}
\sum_{\sg} \f{\sg}{2} \left< J^f_{\bk\sg,\bk'\sg} J^d_{\bm{k',k}} f(\hat \bk, \hat \bk')  \right>_{\Omega},  \label{eq:average1}
\end{\eq}
where $\displaystyle f(\hat \bk, \hat \bk') \equiv \f{m^2}{k^2} \f{\rd \eps_{\bk}}{\rd k_x}\f{\rd \eps_{\bk'}}{\rd k'_y} = \sin \theta_{k} \cos \phi_{k} \sin \theta_{k'} \sin\phi_{k'} $,
and $\left< \cdot \cdot \cdot \right>_{\Omega}$ denotes the average over
the Fermi level, which is defined as $\displaystyle \left< A(\hat \bk_1,\cdot \cdot \cdot, \hat \bk_n) \right>_{\Omega} \equiv \int \f{d\Omega_{k_1} \cdot \cdot \cdot d\Omega_{k_n}}{(4\pi)^n} A(\hat \bk_1 \cdot \cdot \cdot \hat \bk_n) $.
Since $f(\hat \bk, \hat \bk') =2\pi \left\{ Y^1_{1}(\hat \bk) - Y^{-1}_1(\hat \bk) \right\} \left\{ Y^{-1}_{1}(\hat \bk') + Y^{1}_1(\hat \bk') \right\}/3i $, angular integration such as
$\int d\Omega_{k} Y^{M-\sg/2}_{3}(\hat \bk)Y^{m}_{l}(\hat \bk)Y^{\pm1}_{1}(\hat 
\bk)$ appears in eq. (\ref{eq:average1}). This integral is finite only when $l=2,4$.
Therefore, the interference of the $f (l=3)$ and $d \ (l=2)$ partial waves 
is essential for skew scattering \cite{Fert}.
After performing the angular integrations, eq. (\ref{eq:average1})
is given as
\begin{\eq}
{\rm eq}. (\ref{eq:average1}) = i \f{4}{7} J_d J_f,
\nonumber
\end{\eq} 
%
Using the relations $|G^R_{\bk}(0)|^2 \approx \f{\pi}{\g} \delta(\mu-\eps_{\bk})$
for small $\g$, 
and $\f{1}{N}\sum_{\bk_1} G^R_{\bk_1}(0)= g^R(0)= -i \pi N(0)$ ,
eq. (\ref{eq:ss-born1}) is transformed into
\begin{\eq}
\sxy^{\rm ss} &=& -\frac{e}{2\pi } \frac{4 \pi^3 }{7} n_{\rm imp} J_d J^2_f N(0) 
\nn \\
&\times& \sum_{\bk,\bk'} \frac{\partial \varepsilon_k }{\partial k} 
\frac{\partial \varepsilon_k' }{\partial k'} \frac{1}{\g^2}
\delta(\mu-\varepsilon_k) \delta(\mu-\varepsilon_{k'}). \label{eq:ss-born2}
\end{\eq}
Since $\g_f= 3\pi n_{\rm imp} N(0) J_f^2$
in the Born approximation,
$\sxy^{\rm ss}$ is given by
\begin{\eq}
\sigma^{\rm ss}_{\rm SH} = -\frac{e}{2\pi} \frac{1}{21\pi^2} J_d k^4_F \frac{\g_f}{\g^2}. \label{eq:ss-Born}
\end{\eq}

Now, we derive the skew scattering term 
using the $T$-matrix approximation, which gives the exact result for $n_{\rm imp} \ll 1$.
In this case, $T^{f(2) R}_{\bk\sg,\bk'\sg}(0)$ in eq. (\ref{eq:ss-born1}) is replaced with the 
full $T$-matrix
$T^{fR}_{\bk\sg,\bk'\sg}(0)= T^{f(2) R}_{\bk\sg,\bk'\sg}(0) 
\left(1- g^R(0) J_f \right)^{-1}$,
where the first order term in $J^f_{\bk\sg,\bk'\sg}$ has been dropped.
The diagrammatic expression for $\sxy^{\rm ss}$ is shown in Fig. \ref{fig:ss-born} (b).
The angular integration in the $T$-matrix approximation
can be performed as 
\begin{\eq}
& & {\rm Re} \sum_{\sg} \f{\sg}{2} \left< T^{fR}_{\bk\sg,\bk'\sg}(0) J^d_{\bk',\bk}  f(\hat \bk, \hat \bk') \right>_{\Omega}
= \f{4}{7} \f{J_d J^2_f \pi N(0)}{1+ \left( \pi N(0) J_f \right)^2}.
\end{\eq}
Thus, $\sxy^{\rm ss}$ in the $T$-matirx approximation is given by
$\left( 1+ (\pi N(0) J_f)^2 \right)^{-1}$ times eq. (\ref{eq:ss-born2}).
Therefore, 
the expression for the skew scattering term in eq. (\ref{eq:ss-Born}) 
is valid beyond the Born approximation, by considering $\g_f$ given by 
eq. (\ref{eq:gamma-T}).

Here, we discuss the temperature dependence of $\sxy^{\rm ss}$.
From eq. (\ref{eq:ss-Born}), $\sxy^{\rm ss}$ is proportional to $1/\g_f$ when $\g_f \gg \g_0$. Since $\g_f \propto -\ln \left( T/W_{\rm band} \right)$ 
for $T \gg T_{\rm K}$ \cite{Hewson},
$\sxy^{\rm ss}$ decreases as the temperature decreases.
In contrast, $\sxy^{\rm ss} \propto \g^f$ for $\g_f \ll \g_0$,
which increases at low temperatures if $\g_0$ is constant. 

\subsection{ Side Jump Term}

\begin{figure}[!htb]
\includegraphics[width=1.0\linewidth ]{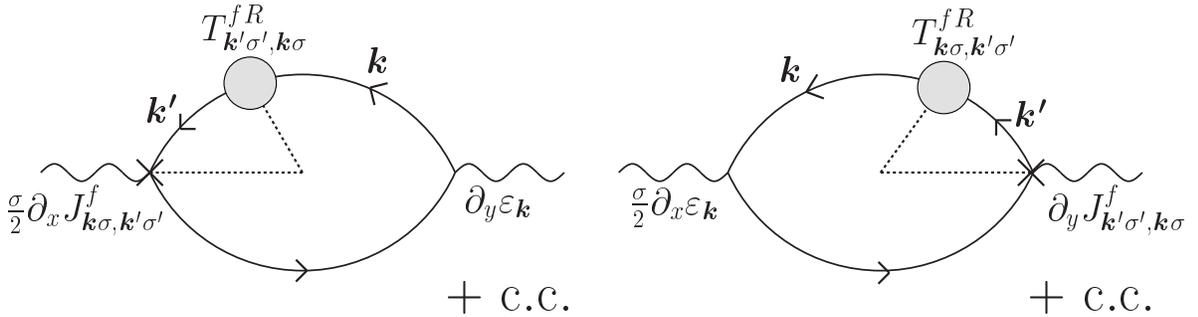}
\caption{\label{fig:sj-born} Diagrammatic expression for the side jump term
in the $T$-matrix approximation. 
Both two diagrams give the same contribution.
Note that $J_d$ is not necessary for the side jump.
} 
\end{figure}

In this section, we briefly discuss the side jump term $\sxy^{\rm sj}$.
In the $T$-matrix approximation, $\sxy^{\rm sj}$ is given by \cite{Bruno}
\begin{\eq}
\sxy^{\rm sj} &=& -\f{e}{2 \pi } \frac{1}{N^2} \sum_{\bk\sg,\bk'\sg'}  \sg \f{\rd J^f_{\bk\sg,\bk'\sg'}}{\rd k_x}  \f{\rd \eps_{\bk}}{\rd k_y} |G^R (0)|^2 \nn \\
&\times& \left[ T^{fR}_{\bk\sg,\bk'\sg'} G^R_{\bk'}(0) + \left< R \leftrightarrow A \right> \right],
\end{\eq}
whose diagrammatic expression is shown in Fig. \ref{fig:sj-born}.
%
%
In contrast to the skew scattering term, $J_d$ is not necessary for the side 
jump term. 
As reported in ref. \cite{Kontani94}, the large anomalous velocity $v^{\rm a}_{\bk M \sg}$, which is not
perpendicular to the Fermi surface, arises from the $\bk$-derivative of the phase factor in the $c$-$f$ mixing potential as follows:
\begin{\eq}
\f{\rd V^f_{\bk M \sg}}{\rd k_x} \ni v^{\rm a}_{\bk M \sg} = -i \left( M-\f{\sg}{2} \right) \f{k_y}{k_x^2 + k_y^2} V^f_{\bk M \sg}.
\end{\eq}
The final expression for the side jump term in the $T$-matrix 
approximation is given by
\begin{\eq}
\sxy^{\rm sj} = -\f{e}{2\pi} \f{4}{3} \f{k_F}{\pi} \f{\g_f}{\g}, \label{eq:sj-tmat}
\end{\eq}
where we have used the relations 
$\sum_{M \sg} M \sg |V^f_{\bk M\sg}|^2 = 3|V_f|^2 (1-\sin^2 \theta) $ and 
$\sum_{M\sg} \sg^2 |V^f_{\bk M\sg}|^2 = 6 |V_f|^2$.
By using eqs. (\ref{eq:ss-Born}) and (\ref{eq:sj-tmat}), the ratio 
$\sxy^{\rm sj}/\sxy^{\rm ss}$ is given by $\displaystyle\f{28}{3\pi} \f{\gamma}{J_d 
n}$. 
Therefore, $\sxy^{\rm sj}$ 
will exceed $\sxy^{\rm ss}$ in dirty metals.
However, the skew scattering contribution will be dominant in simple metals
with magnetic impurities for $n_{\rm imp} \simle 0.01 \ (\rho_0 \simle 10\mu \Omega$cm) \cite{Fert}.

\section{Discussions}

\subsection{Origin of the Skew Scattering Mechanism}

In section \ref{SS-term}, we have studied the SHC due to skew scattering by using the 
Green function method. In this section, we discuss the origin of the skew scattering mechanism
based on the Boltzmann transport theory.
For this purpose, we study a simplified two-orbital model with $M=\pm 5/2$,
assuming the strong crystalline electric field.
In this model, $V^f_{\bk M \sg} =  -\sg \sqrt{4\pi/7}V_f \left\{ \sqrt{6} Y^{-3\sg}_{3} (\hat \bk) \delta_{M,-5/2\sg} + Y^{2\sg}_{3} (\hat \bk) \delta_{M,5/2\sg} \right\}$. 
Since $V^f_{\bk M \sg} \propto Y^{-3\sigma }_3 (\hat \bk) \propto {\rm e}^{-3 i \sg \phi_{k}}$ approximately,
the second order term of $T^f$ is simply given as
$T^{f (2) R} \sim -i N(0) {\rm e}^{-3i \sg 
\left( \phi_{\bk}-\phi_{\bk'} \right)} $.  
In the Boltzmann transport theory, the spin Hall resistivity due to skew 
scattering is
$\rho^{\rm ss}_{\rm SH} \propto \sum_{\sg} \sg \left< k_x k'_y w(\bk\sg \rightarrow \bk'\sg) \right>_{\rm FS}$, where 
$w$ represents the scattering probability,
which is proportional to $ \displaystyle n_{\rm imp} \left|T^{f(2) R}_{\bk\sg,\bk'\sg} + J^d_{\bk,\bk'} \right|^2$ due to Fermi's golden rule in the present model.
According to ref. \cite{Fert},
skew scattering occurs when the scattering probability includes
an assymetric component 
$w^{\rm ss}(\bk \rightarrow \bk') \propto {\rm Im} {\rm e}^{i(\phi_k -\phi_{k'})} \propto (\hat \bk \times \hat \bk')_z$.
In the present model, $w^{\rm ss}$ arises from
the interference of the $f$ and $d$ scattering channel, 
$w^{\rm ss} \in \left( T^{f(2)R}_{\bk\sg,\bk'\sg} J^d_{\bk,\bk'} + {\rm c.c} \right)$.
In fact, $ w^{\rm ss}(\bk\sg \rightarrow \bk'\sg) \propto {\rm Im} {\rm e}^{i\sg (\phi_k - \phi_{k'})} $ in this model 
since $J^d_{\bk,\bk'}$ contains the term
$Y^{\pm 2}_{2} (\hat \bk) \left[ Y^{\pm 2}_{2}(\hat \bk')  \right]^{\ast}  \propto {\rm e}^{\pm 2i (\phi_{k}-\phi_{k'})} $. 
In summary, a conduction electron with $\sg$ hybridizes with $l_z=-3\sg$
state due to the strong SOI, and therefore the spin-dependent skew scattering 
probability $w^{\rm ss} (\bk\sigma \rightarrow \bk'\sigma) \propto {\rm Im} {\rm e}^{i\sg (\phi_k -\phi_{k'})}$ 
arises from the interference of the $f$ and $d$ angular momenta.
Thus, the origin of the SHE due to skew scattering mechanism is well understood
based on the simplified two-orbital model.

\subsection{Estimations of the Spin Hall angle and the Skew Scattering Term}

First, we estimate the magnitude of the spin Hall angle 
$\displaystyle \tan\aH \equiv \f{\sxy^{\rm ss}}{\sg_{xx}}\f{2e}{\hbar} $.
The longitudinal conductivity is given by
$\displaystyle \sg_{xx} = \f{e^2 n}{2 m \g}$,
where $n=k^3_F /3\pi^2$ is the density of the
conduction electrons.
Then, the spin Hall angle is given by
\begin{\eq}
\tan \alpha_{\rm H} = \f{8\pi^2}{7} J_d N(0) \f{\g_f}{\g}. \label{eq:Hallangle}
\end{\eq}
%
Since 
$\tan \delta_2 = -\pi N(0) J_d$ \cite{Hewson},
eq. (\ref{eq:Hallangle}) can be 
rewritten as $\displaystyle \tan \aH = -\f{8\pi}{7}\f{\g_f}{\g} \delta_2$ if $|J_d| N(0) \ll 1$.
In the case where $\g_f \gg \g_0$, $\aH$ reaches $\sim 0.3$ for $\delta_2 \sim 0.1$ \cite{AgYb}.
Therefore, giant SHE will be realized in metals with rare-earth impurities.


Recently, giant SHE was observed in FePt/Au devices \cite{Seki}.
It was reported that the large spin Hall angle in Au
is $\sim 0.1$.
Since the intrinsic SHE in Au is small 
due to the Fermi surface being mainly composed of $s$-electrons 
\cite{Tanaka-4d5d}, the observed giant SHE may 
be explained by the skew scattering mechanism.
Very recently, the extrinsic SHE in AuFe metal has been studied  
with the first-principles band structure calculations 
independently of our study \cite{Nagaosa-FeAu}.

Second, we estimate the magnitude of the skew scattering term
for $\delta_2 =0.1$, where the spin Hall angle is $0.3$.
Then, $\sigma^{\rm ss}_{\rm SH}= (\hbar/2e) 0.3 \times \sigma_{xx}$ from the 
above discussion.
In the case where the resistivity is $10 \mu\Omega$cm,
we obtain $\sigma^{\rm ss}_{\rm SH} = 15000 (\hbar/e) \Omega^{-1}$cm$^{-1}$
if we put the length of unit cell $a = 4 {\rm \AA}$, which is the case for Pt and Au.
The obtained skew scattering term is about $60$ times larger than
the SHC observed in Pt: $240 (\hbar/e) \Omega^{-1}$cm$^{-1}$.
Since the resistivity in Pt and Au is $\rho \sim 10\mu\Omega$cm and $\rho \sim 1 \mu 
\Omega$cm, respectively, the skew scattering term due to magnetic impurities 
should dominate the intrinsic term.
In contrast, the skew scattering due to ``natural randomness" has not been observed 
in the AHE in transition metals, even in $\rho \sim 10\mu\Omega$cm \cite{Asamitsu}.
Therefore, the orbital degrees of freedom of the impurity is indispensable for 
the giant extrinsic SHE.


Finally, we compare the magnitude of $\sigma^{\rm ss}_{\rm SH}$ and $\sigma^{\rm sj}_{SH}$.
If we put $k_{F}=\pi/a$ and $\gamma \approx \gamma_f$ in eq. (\ref{eq:sj-tmat}),
$\sigma^{\rm sj}_{\rm SH} = \f{-e}{2\pi a} \f{4}{3} \sim -(4/3) \times 10^{3} (\hbar/e)\Omega^{-1}$cm$^{-1}$ for $a = 4 {\rm \AA}$.
Therefore, the skew scattering term will be comparable in magnitude to
the side jump term when $\rho \sim 100 \mu\Omega$cm.
Thus, the skew scattering term is dominant over the side jump term when 
$ \rho \ll 100\mu\Omega $cm.



\section{Summary}
In summary, we have studied the extrinsic SHE due to magnetic impurities,
where Ce and Yb impurities have been discussed as a typical case. 
The obtained expression for the skew scattering term shows that
the spin Hall angle 
reaches $O(10^{-1})$, which is more than 10 times larger than that in Pt.
Therefore, we propose that the extrinsic SHE due to rare-earth impurities is useful as
a method for creating large spin current efficiently.
The present study also suggests that the giant
SHE will arise from other rare-earth atoms as well as 4$d$ and 5$d$ atoms
that possesses strong SOI.
As in the case for the intrinsic SHE, orbital degrees of freedom are 
essential for generating a huge extrinsic SHE.
According to eq. (\ref{eq:Hallangle}), giant spin Hall angle always appears 
when $\g_f \gg \g_0 $. 
Thus, the Kondo resonance, $N(0) J_f \gg 1$, is not a necessary 
condition for the giant SHE. 
In contrast, the AHC ($=\sigma_{xy}/H_z$) is enhanced by the Kondo effect
since it is proportional to the uniform susceptibility.

In a similar way to the present study, we can obtain $\sxy^{\rm ss}$ and $\sxy^{\rm sj}$ for $J=7/2$, which 
is the case for Yb-impurities.
In this case, $a^M_{m\sg}= \{(7/2 + M\sg)/7 \}^{1/2} \delta_{m,M-\sg/2}$, and
$\g_f$ is given by $(4/3)$ times eq. (\ref{eq:gamma-T}).
The obtained $\sxy^{\rm ss}$ and $\sxy^{\rm sj}$ are given by 
$(-3/4)$ times eq. (\ref{eq:ss-Born}) and eq. (\ref{eq:sj-tmat}), respectively.


%

\section*{Acknowledgements}
We are grateful to D. S. Hirashima and J. Goryo for fruitful discussions.


\section*{References}
\begin{thebibliography}{10}

\bibitem{Sinova-SHE}
J. Sinova, D. Culcer, Q. Niu, N. A. Sinitsyn, T. Jungwirth, and 
A. H. MacDonald,
Phys. Rev. Lett. {\bf 92} (2004) 126603.

\bibitem{Murakami-SHE}
S. Murakami, N. Nagaosa and S.C. Zhang,
Phys. Rev. B {\bf 69} (2004) 235206.

\bibitem{Saitoh} 
E. Saitoh, M. Ueda, H. Miyajima and G. Tatara,
Appl. Phys. Lett. {\bf 88} (2006) 182509.

\bibitem{Kimura}
T. Kimura, Y. Otani, T. Sato, S. Takahashi, and S. Maekawa,
Phys. Rev. Lett. {\bf 98} (2007) 156601;
L. Vila , T. Kimura, and Y. Otani,
Phys. Rev. Lett. {\bf 99}, 226604 (2007); 
Y. Otani et al., (unpublished)

\bibitem{Kontani-Ru} 
H. Kontani, T. Tanaka, D. S. Hirashima, K. Yamada, and J. Inoue,
Phys. Rev. Lett. {\bf 100}, 096601 (2008).

\bibitem{Kontani-Pt}
H. Kontani, M. Naito, D. S. Hirashima, K. Yamada, and J. Inoue,
J. Phys. Soc. Jpn. {\bf 76} 103702 (2007).


\bibitem{Tanaka-4d5d}
T. Tanaka, H. Kontani, M. Naito, T. Naito, D. S. Hirashima, K. Yamada, and J. Inoue,
Phys. Rev. B {\bf 77}, 165117 (2008).

\bibitem{Kontani-OHE}
H. Kontani, T. Tanaka, D. S. Hirashima, K. Yamada, and J. Inoue, 
arXiv:0806.0210.

\bibitem{Guo-Pt}
G. Y. Guo, S. Murakami, T.-W. Chen, and N. Nagaosa, Phys. Rev. Lett. {\bf 100}, 096401 (2008).

\bibitem{Dyakonov}
M. I. Dyakonov and V. I. Perel, JETP Lett. {\bf 13} 467 (1971).

\bibitem{Hirsh}
J. E. Hirsch,
Phys. Rev. Lett. {\bf 83}, 1834 (1999).

\bibitem{Takahashi}
S. Takahashi and S. Maekawa, 
J. Phys. Soc. Jpn. {\bf 77}, 031009 (2008).

\bibitem{Smit}
J. Smit, Physica {\bf 24} (1958) 39.

\bibitem{Berger}
L. Berger, Phys. Rev. B {\bf 2} (1970) 4559.

\bibitem{Bruno}
A. Crepieux and P. Bruno, Phys. Rev. B {\bf 64} 014416 (2001).

\bibitem{Fert}
A. Fert, J. Phys. F {\bf 3}, 2126 (1973), 
A. Fert and A. Friederich,
Phys. Rev. B {\bf 13}, 397 (1976).


\bibitem{Fert-Jaoul}
A. Fert and O. Jaoul, Phys. Rev. Lett. {\bf 28}, 303 (1972).


\bibitem{Hewson}
A. C. Hewson, $\textit{The Kondo Problem to Heavy Fermions }$ 
(Cambrige Studies in Magnetism, 1993);
K. Yosida, $\textit{Theory of Magnetism}$
(Springer 1996).

\bibitem{comment}
We omit the atomic SOI for $d$-electrons since
it is much smaller than $|\mu -E^d|$.




\bibitem{YY}
K. Yamada and K. Yosida: Prog. Theor. Phys. {\bf 53} (1975) 970.

\bibitem{Kontani94}
H. Kontani and K. Yamada, J. Phys. Soc. Jpn. {\bf 63}, 2627 (1994).

\bibitem{Haldane}
F. D. M. Haldane,
Phys. Rev. Lett. {\bf 40}, 416 (1978).

\bibitem{Seki}
T. Seki, Y. Hasegawa, S. Mitani, S. Takahashi, H. Imamura, S. Maekawa, J. Nitta, and K. Takanashi,
Nature {\bf 7}, 125 (2008).

\bibitem{AgYb}
Recently first-principles band structure calculation has been performed by 
the present authors,
and $\delta_2$ is estimated as $\sim 0.1$ for Yb impurities in Ag host.

\bibitem{Nagaosa-FeAu}
G. Y. Guo, S. Maekawa, and N. Nagaosa, arXiv:0809.2921.

\bibitem{Asamitsu}
T. Miyasato, N. Abe, T. Fujii, A. Asamitsu, S. Onoda, Y. Onose, N. Nagaosa, and Y. 
Tokura, 
Phys. Rev. Lett. {\bf 99} (2007) 086602.


\end{thebibliography}
\end{document}